\documentclass[usenatbib,usegraphicx]{mn2e}
\usepackage{color}
\begin{document}

\title[Data processing and lightcurve production]
  {The Monitor project: Data processing and lightcurve production}
\author[J.~M.~Irwin et al.]{Jonathan~Irwin$^{1}$, Mike~Irwin$^{1}$,
  Suzanne~Aigrain$^{1}$, Simon~Hodgkin$^{1}$,
\newauthor
Leslie~Hebb$^{2}$, Estelle~Moraux$^{3}$ \\
$^{1}$Institute of Astronomy, University of Cambridge, Madingley Road,
  Cambridge, CB3 0HA, United Kingdom \\
$^{2}$School of Physics and Astronomy, University of St Andrews,
  North Haugh, St Andrews, KY16 9SS, Scotland \\
$^{3}$Laboratoire d'Astrophysique, Observatoire de Grenoble, BP 53,
  F-38041 Grenoble C\'{e}dex 9, France}

\maketitle

\begin{abstract}
We have begun a large-scale photometric survey of nearby open clusters
and star-forming regions, the Monitor project, aiming to measure
time-series photometry for $> 10,000$ cluster members over
$> 10\ {\rm deg^2}$ of sky, to find low-mass eclipsing binary and
planet systems.  We describe the software pipeline we have developed
for this project, showing that we can achieve peak RMS accuracy over
the entire data-set of better than $\sim 2\ {\rm mmag}$ using aperture
photometry, with RMS $< 1 \%$ over $\sim 4\ {\rm mag}$, in data from
$2$ and $4\ {\rm m}$ class telescopes with wide-field mosaic cameras.
We investigate the noise properties of our data, finding correlated
`red' noise at the $\sim 1-1.5\ {\rm mmag}$ level in bright stars,
over transit-like timescales of $2.5\ {\rm hours}$.  An important
source of correlated noise in aperture photometry is image blending,
which produces variations correlated with the seeing.  We present a
simple blend index based on fitting polynomials to these variations,
and find that subtracting the fit from the data provides a method to
reduce their amplitude, in lieu of using techniques such as point
spread function fitting photometry which tackle their cause.  Finally,
we use the {\sc Sysrem} algorithm to search for any further systematic
effects.
\end{abstract}

\section{Introduction}
\label{intro_section}

The Monitor project is a large-scale photometric survey of galactic
open clusters and star forming regions.  We intend to measure
high-cadence time series photometry for $> 10,000$ cluster
members over $> 10\ {\rm deg^2}$ of sky, aiming to find
the first transiting planets in open clusters, and tens-hundreds of
low-mass eclipsing binary systems, possibly including brown dwarfs.
For more details of the project's scientific goals and the results of 
simulations giving likely numbers of detected systems, the reader is
referred to \citet{a2006}, hereafter paper I.  A brief summary of the
project is also given in \citet{hodg06}.

Data processing in this project is challenging.  In a typical night we
obtain $\sim 25\ {\rm Gigabytes}$ of imaging data using the Wide Field
Camera (WFC) on the Isaac Newton Telescope (INT), and this can be as
large as $\sim 50\ {\rm Gigabytes}$ for some of the other instruments
we are using (for example MegaCam on the Canada-France-Hawaii
Telescope, hereafter CFHT).  Since our survey covers $9$ clusters over
$> 10$ nights per cluster, this is a multi-terabyte project.

\citet{kf92} give a detailed discussion of differential photometry
problems and techniques, from the point of view of attempting to
detect low-amplitude stellar oscillations, but many of their arguments
apply equally to transit surveys.  Using CCD cameras, one can perform
differential photometry on very large numbers of stars simultaneously,
using non-variable stars in the field as comparison sources to remove
transparency (and other) variations in the atmosphere.  Differential
photometric precision at the sub-$1\%$ level can be readily achieved
using this method, even in somewhat non-photometric conditions.

Our methodology is based on experience gained by members of our group
from the University of New South Wales Extrasolar Planet Survey
\citep{h2005}, and much of the pipeline code is now shared between the
two projects.

We describe the observations in \S \ref{obs_section} and the basic CCD
data reduction in \S \ref{dr_section}.  \S \ref{dp_section} gives an
overview of the steps required to produce differential photometry, and
hence lightcurves, from these data, and  the practical details of
their implementation are discussed in \S \ref{implement_section} and
\S \ref{lc_section}.

In \S \ref{noise_section} we examine the noise properties of our data,
with particular attention given to correlated (`red') noise, which can
be a serious problem in differential photometry \citep*{pzq06}.  \S
\ref{seeing_section} examines one particular source of correlated
noise, namely seeing-correlated variations induced in the lightcurves
by blending of flux from neighbouring sources into the photometric
apertures, and in \S \ref{sysrem_section} we apply the {\sc Sysrem}
algorithm \citep*{t2005} to search for any further sources of
correlated noise in the data.  Finally, we summarise our conclusions
in \S \ref{conc_section}.

\section{Observations}
\label{obs_section}

We are using wide-field mosaic cameras on several telescopes to
perform the survey, principally: the Wide Field Camera (WFC) on the
$2.5\ {\rm m}$ INT ($4\times$ 2k$\times$4k CCDs, $\sim 34\times34'$ 
field-of-view) and MegaCam on the $3.5\ {\rm m}$ CFHT ($36\times$
2k$\times$4.5k CCDs, $\sim 1\times1^\circ$ FoV) in the Northern
hemisphere, and the ESO/MPG $2.2\ {\rm m}$ Wide Field Imager (WFI)
($8\times$ 2k$\times$4k CCDs, $\sim 34\times33'$ FoV) and Mosaic II on
the $4\ {\rm m}$ CTIO Blanco telescope ($8\times$ 2k$\times$4k CCDs,
$\sim 37\times37'$ FoV) in the Southern hemisphere.  Due to the
enormous quantity of data, a uniform strategy for observing
(where possible) and data processing is essential.

The peculiarities of scheduling for each of these telescopes limit
our flexibility in observing strategy so this will be discussed only
briefly.  We observe in $i'$ or $I$, since this maximises
signal-to-noise for our faint, red objects of interest, and minimises
any colour-dependent atmospheric extinction, which can be difficult to
correct in the lightcurves.  The wide-field mosaic instruments we are
using typically suffer from fringing in red bandpasses, so the
SDSS-like $i'$ filter \citep{f96} is preferred where available, since
this minimises fringing due to its sharp red cut-off at $\sim 8500$
\AA, compared to the long red tail of the standard $I$ filters.

Exposure times are selected to give good signal to noise on the
largest possible number of cluster members, while keeping the targets
sufficiently bright that medium-resolution follow-up observations on
$4\ {\rm m}$ class telescopes and high-precision radial velocities on
$8\ {\rm m}$ class telescopes remain feasible.  Typically our
exposures are in the range $30 - 120\ {\rm s}$, so the survey
efficiency is overhead dominated with the slow readout times for the
mosaic instruments we are using (most are $\sim 60\ {\rm s}$).  In
several cases we cycle between multiple fields in a single cluster to
increase our spatial coverage, or between multiple clusters, but we
aim to obtain an observing cadence no worse than $15\ {\rm minutes}$
for clusters where we are primarily searching for eclipsing binaries,
and $5\ {\rm minutes}$ for planet searches, or where short-term
stellar variability is a problem, ie. the youngest clusters (see
paper I for more details).

Accurate flat fielding is of critical importance in differential
photometry, so we take extra care to ensure that this is done as well
as possible.  We find that twilight flat fields provide superior
results compared to dome flat fields for all the instruments we are
using, provided sufficient signal can be accumulated.  For a typical
detector with gain of a few ${\rm e^-/ADU}$, and a typical twilight
flat illumination level of $20,000\ {\rm ADU/pixel} = 40,000\ {\rm
  e^-/pixel}$, the Poisson noise is $200\ {\rm e^-}$, ie. a
signal-to-noise ratio of $200$, which is equivalent to $\sim 5\ {\rm
  mmag}$ photon noise per pixel.  Averaged over a typical photometric
aperture of $3\ {\rm pixel}$ radius this gives $\sim 1\ {\rm mmag}$ --
ie. a significant contribution.  Over a typical one week observing
run, we can readily obtain at least $25$ flat field frames, which
reduces the Poisson noise to $\sim 0.2\ {\rm mmag}$, a level which
is perfectly acceptable for our purposes. 

A related issue is that of positioning the telescope.  Even using the
flat fielding procedure described, small errors of the order of $0.1 -
1 \%$ remain in the flat field frames, and fringing in the detectors,
even after correction, can reach amplitudes of $\sim 0.2\ \%$.
The effects can be divided into low spatial frequencies, dominated by
non-uniform illumination of the flat field frame, and high spatial
frequencies, eg. fringing, or differential variations in the
quantum efficiency of the pixels (eg. as a function of wavelength,
since the spectra of the flat field source and target star are
different).  The combination of these 
effects typically limits the achievable photometric precision to a few
mmag depending on the instrument, in our experience.  In order to
minimise these effects we therefore aim to reposition each star on {\it
  exactly} the same pixel of the detector in each exposure.  This is
done by using the telescope guiding system to correct for pointing
errors, where available.  We note in passing that this procedure may
introduce correlated noise (see \S \ref{noise_section}), particularly
in the event that any positioning errors are periodic or result in a
slow drift across a few pixels of the detector.  It has been suggested
that an intentional random jitter in the telescope positions may prove
beneficial to convert this source of correlated noise to a source of
random noise.  However, due to the need to move over a larger region
of the detector, doing this is likely to introduce greater effects due
to flat fielding errors, fringing, and other effects operating over
short spatial scales.  It therefore carries an inherent risk of
raising the overall noise level, and thus would require more data, so
we have been unable to explore it further as telescope time is always
at a premium when using large international facilities.

Equatorial standard star fields (from the catalogue of \citealt{l92})
are observed regularly during our observing runs, to provide
calibrated photometry on a standard zero-point system.

\section{Data reduction}
\label{dr_section}

The need for a uniform data processing strategy was highlighted in \S
\ref{obs_section}.  We employ a modified version of the INT/WFC data
reduction pipeline, developed for the INT Wide Field Survey (WFS) and
originally described in \citet{il2001}.  This has been
successfully applied to data from all of the instruments mentioned in
\S \ref{obs_section} at the time of writing.

\begin{figure}
\centering
\includegraphics[angle=270,width=3.2in]{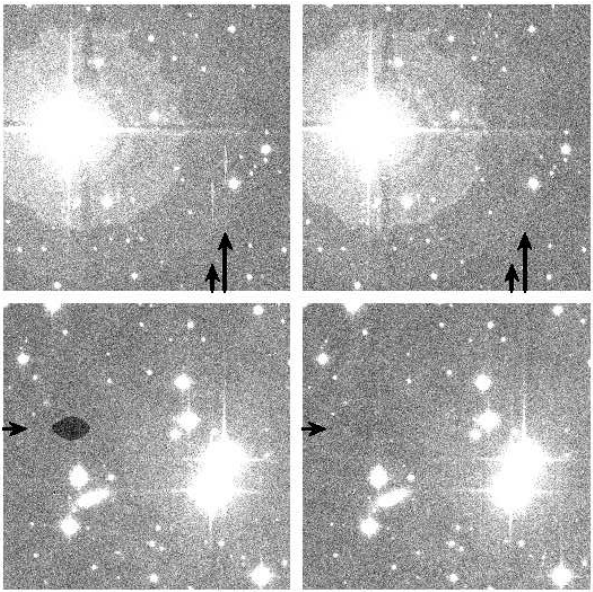}

\caption{A section of CCDs 3 (top) and 4 (bottom) of an INT/WFC image
of M34, before (left) and after (right) applying the cross-talk
correction described in the text.  The images show positive cross-talk
($f_{ij} > 0$) from CCD 4 to CCD 3 (eg. at the position in the top
panel corresponding to the pair of bright stars visible at the right
of the bottom panel, marked with arrows), and negative cross-talk
($f_{ij} < 0$) from CCD 3 to CCD 4 (eg. at the position in the bottom
panel corresponding to the brightest star on the left hand side of the
top panel, marked with an arrow).}

\label{m34_crosstalk}
\end{figure}

Two of the instruments we are using (INT/WFC and CTIO Mosaic) suffer
from electrical cross-talk between the detector readouts, the effect
of which is illustrated in Figure \ref{m34_crosstalk}.  For the
INT/WFC the maximum level is $\sim 4 \times 10^{-4}$, typically a 
sufficiently low level to be ignored, but for the CTIO Mosaic the
level is $\sim 2 \times 10^{-3}$.  Therefore, before starting the
standard CCD reduction procedure this must be corrected, and is done
in a simple manner by subtracting a fraction $f_{ij}$ of the detected
counts on detector $i$ from detector $j$.

We then follow the standard CCD reduction scheme of bias correction,
trimming of overscan and non-illuminated regions, non-linearity
correction, flatfielding and gain correction, followed by defringing,
catalogue generation, astrometric and photometric calibration
described in \citet{il2001}.  We use the point source catalogue
(PSC) from the Two-Micron All Sky Survey (2MASS) as an astrometric
reference catalogue, which we find gives typical RMS residuals of $<
0.1''$. 

\section{Differential photometry}
\label{dp_section}

In the discussion that follows, we use aperture photometry.  The
technique we use is similar to standard aperture photometry, except
our apertures are `soft-edged', and overlapping sources are fitted
simultaneously using circular top-hat functions as the `PSF'.  We have
found that for our open cluster fields, this technique is sufficient
to obtain a photometric precision of $\sim 1-2\ {\rm mmag}$ for the
brightest stars, without need to invoke more exotic techniques such as
point spread function fitting (PSF-fitting; eg. \citealt{s87}) or
difference image analysis (DIA; \citealt{al98}, \citealt{a2000}),
although these are discussed briefly in \S \ref{seeing_section}.

\subsection{Background estimation}

Robust, repeatable background estimation is of vital importance in
aperture photometry.  We use a variant of the technique discussed in
\citet{i85} for background estimation in our aperture photometry,
which has been found empirically to work at least as well as the
standard technique of using an annulus around the photometric
aperture, for fields with slowly-varying sky backgrounds.  A brief
description of the method is given here, and the reader is referred to
\citet{i85}, \citet{i96} and \citet{mji06} for a more detailed discussion.

Briefly, the image is divided into a coarse grid of $64 \times 64\
{\rm pixel}$ bins ($\sim 20\ {\rm arcsec}$ on sky).  The background
level in each bin is estimated using a robust $k\sigma$ clipped median
of the counts in that bin, using the robust median of absolute
deviations (MAD; eg. \citealt*{hoag83}) estimator to calculate $\sigma$,
and rejecting bad pixels using the confidence maps (see
\citealt{il2001}).  The resulting map is filtered using 2-D bilinear
and median filters to avoid problems due to single bins dominated by
bright stars.  The background in a given image pixel can then be
estimated using bilinear interpolation over the coarse background map.

\subsection{Aperture placement}

Differential photometry is very sensitive to small positioning errors
when placing photometric apertures on the science images.  For a
Gaussian PSF, the error in the derived fluxes is given to first order
by:
\begin{equation}
{\delta F\over{F}} \approx {1\over{\sqrt{2\pi}}}\ {\delta
  x\over{\sigma}}\ {2 r \delta x \over{\sigma^2}}\ {\rm e}^{-r^2/2\sigma^2}
\label{applac_equation}
\end{equation}
where $\delta x$ is the positioning error, $r$ is the radius of the
aperture, and $\sigma$ describes the PSF size (ie. seeing, ${\rm FWHM}
\approx 2.35 \sigma$).  See Appendix \ref{applac_deriv} for a derivation.

Typically we set $r = 2.35 \sigma$, ie. an aperture radius equal to
the image FWHM, so:
\begin{equation}
{\delta F\over{F}} \sim 0.119\ {\delta x^2\over{\sigma^2}}
\end{equation}
Taking for example a typical value $\delta x = 0.1 \sigma$, this implies a
flux error of $\approx 1\ {\rm mmag}$.  Eq. (\ref{applac_equation})
also confirms the intuitive result that using a larger aperture
reduces the effect of centroid errors, at the cost of increased noise
from the sky background.

We therefore first consider the question of how best to determine the
correct locations for the apertures.

The `default' technique used by existing source extraction software,
as included in our pipeline (\citealt{i85}, \citealt{il2001}), or
SExtractor \citep{ba96}, is to find the centroid of each star on the
CCD frame in question, to place an aperture at this position, and
measure the flux.  The accuracy to which this can be done for a star
measured with signal to noise ratio $S$ improves in proportion to $1 /
S$ (eg. \citealt{i85}), giving the general `rule of thumb' that the
error in the image centroid is $\Delta x / S$ where $\Delta x$ is the
sampling interval (pixel scale), implying in general a decrease in the
accuracy of aperture placement moving to fainter stars.

A further problem is that as the seeing changes, the amount of
blending in very close sources will also vary, to the point that they
could become resolved in frames with good seeing, and unresolved in
frames with poor seeing.  This causes the centroid to shift in the
unresolved (or poorer seeing) image toward the companion star, and
hence results in a serious error in the aperture flux measurements.

The standard method for solving these problems, which we call
`co-located aperture photometry', is therefore to use as many stars as
possible to determine the aperture positions, in two stages.  The
first is to determine accurately the {\it relative} centroid positions
of all the stars on the frame, which will be the same for all frames
in the time-series (provided the stars do not move).  This can be done
using a stacked image to increase $S$ (we typically stack the $20$
frames with the best seeing, providing a $\sim$ four-fold improvement
in $S$ over a single frame) and thus obtain an improved master
catalogue with more accurate relative positions.  Furthermore, since
the placement of the apertures remains consistent, the effects of
varying seeing are limited purely to varying flux loss from the
apertures, which can be corrected to a good approximation by a global
normalisation over the frame.

In the second stage, a transformation is computed between this
{\it master frame} and each frame in the time-series on a per-detector
basis, using a standard 6-coefficient linear transformation, derived
using a least-squares fit to a large number of bright stars.  In this
case, the error for the bright stars is dominated by the error in the
transformation, and assuming sufficiently large numbers of stars were
used, this is in turn dominated by errors in the model, eg. due to
radial distortions or other similar effects.  Moreover, any errors in
the mapping from the master frame to the individual frames will
typically either affect all stars in the same way, or will be a
smoothly-varying function of position.  Such effects are readily
removed using a simple polynomial fit (see \S \ref{poly_section}).

\begin{figure}
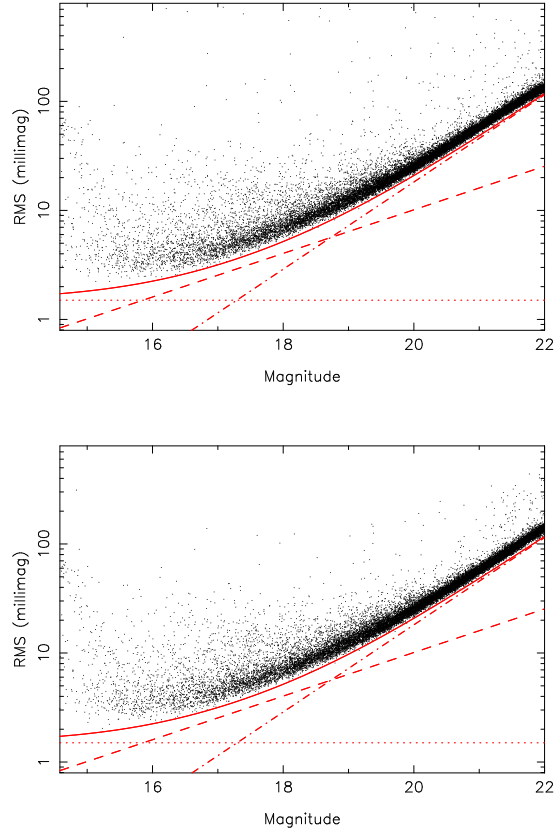

\centering
\includegraphics[angle=270,width=3in]{m50_rms_cent.ps}
\includegraphics[angle=270,width=3in]{m50_rms_cent_comp.ps}

\caption{Plots of RMS scatter as a function of magnitude for the
$i'$-band observations of M50, showing all objects of stellar
morphological classification.  The upper plot shows the results
obtained by placing the photometric apertures at the centroid
positions of the stars, as determined on each frame, and the lower
plot shows the same using the `co-located apertures' technique.  The
plots have been truncated at $i' \sim 22$ since we require the sources
to be detected in at least $10 \%$ of the images for the upper diagram,
and the detections start to become substantially incomplete for fainter
magnitudes.
In both cases, a simple zero point correction of the individual frames
to the master frame has been used (see \S \ref{poly_section}).  The
diagonal dashed lines show the expected RMS from Poisson noise the
object, the diagonal dot-dashed lines show the RMS from sky noise in
the photometric aperture, and the dotted lines show an additional
$1.5\ {\rm mmag}$ contribution added in quadrature to account for
presumed systematic effects.  The solid lines show the overall
predicted RMS, combining these contributions.}

\label{cent_comp}
\end{figure}

Figure \ref{cent_comp} illustrates this for our M50 data.  In this
case we have used a simple constant multiplier to normalise each frame
to the photometric system of the master frame, using an iterative $k
\sigma$ clipped fit (derived from the objects classified as stellar)
to remove any variable stars.  In Monitor data, although there is
little to no improvement using the `co-located apertures' technique
for the majority of sources, it is still necessary to eliminate the
problem of centroid shifts in blended sources, as we have
suggested. We suspect that this is the origin of the spurious variable
sources seen in the upper panel of Figure \ref{cent_comp}. 
Furthermore, another advantage is clear at the faint end, where it
provides much more complete sampling, since we can still place an
aperture and measure the flux even if the object does not pass the
detection threshold on that particular frame, whereas in the upper
panel, the object must be detected and the centroid computed
before this can be done.

For under-sampled data, the required fractional accuracy relative to
the pixel scale is much more stringent, and the noise-induced centroid
errors alone can become highly significant, eg. giving a $\sim 50 \%$
improvement in RMS scatter for significantly under-sampled data from
the University of New South Wales extrasolar planet search
\citep{h2005}.

\subsection{Aperture sizes}

It is straightforward to show that for the majority of images, an
aperture with radius approximately equal to the FWHM of the stellar
images achieves the optimal balance between flux loss (and
consequently, increased Poisson noise in the counts) and integrated
noise in the sky background (which increases with the area of the
aperture).  However, for bright sources, this wastes flux since the
relative size of the sky noise contribution is much smaller, and a
much larger aperture can be used.

Our aperture photometry procedure computes the flux in a sequence of
apertures of radii $r_{\rm core}$, $\sqrt{2}\ r_{\rm core}$,
$2\ r_{\rm core}$, etc. (doubling the area each time) where the `core
radius' $r_{\rm core}$ is set equal to the typical FWHM of stellar
images (and kept fixed for all the data).  We use $r_{\rm core} = 4\
{\rm pixels}$ ($\sim 1.1\ {\rm arcsec}$) for the CTIO-4m+Mosaic data.

We employ a simple procedure to make use of these measurements.  The
lightcurve is computed for each aperture separately, and the root
mean square scatter (computed using a robust median-based estimator)
compared for each source.  We simply choose the aperture with the
smallest RMS for that star.\footnote{The RMS is not an optimal
diagnostic of lightcurve quality for specific purposes
(eg. searching for eclipses, or rotational modulations), since it
reflects the overall scatter rather than, for example, the
correlations in the lightcurve due to systematics.  It is, however,
general-purpose, and thus well-suited for generating lightcurves to
which a wide variety of analysis methods will be applied, as is the
case for the Monitor project.}  This procedure ensures that larger
apertures are used where they give an improvement for bright sources,
but also accounts for blending, where using a larger aperture results
in increased contamination of the flux measurement by neighbouring
stars, and introduces modulations into the lightcurve as the seeing
(and hence the amount of contaminating flux in the aperture) changes.

In order to place all the stars onto the same zero point system, this
procedure necessitates using aperture corrections, to account for the
differing amounts of flux lost from the different sized apertures.
These are computed as simple ratios of the flux measured in the
different apertures, for non-variable stars.

\begin{figure}
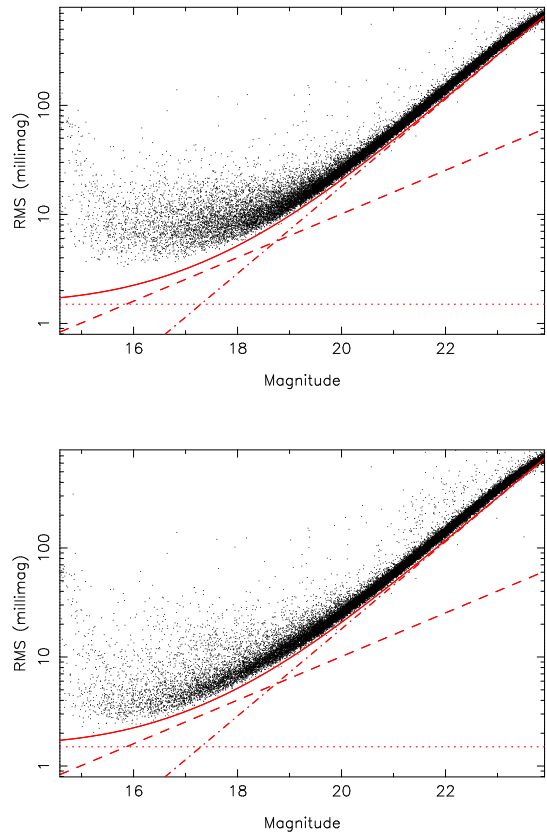

\centering
\includegraphics[angle=270,width=3in]{m50_rms_noaper.ps}
\includegraphics[angle=270,width=3in]{m50_rms_const.ps}

\caption{Plots of RMS scatter as a function of magnitude for the
$i'$-band observations of M50, showing all objects of stellar
morphological classification.  The upper plot shows the results
obtained using a single photometric aperture (radius $r_{\rm core} =
4\ {\rm pixels}$), and the lower plot shows the same using multiple
apertures, selected on a per-star basis.  Lines as Figure
\ref{cent_comp}.  In both cases, a simple zero point correction of the
individual frames to the master frame has been used (see \S
\ref{poly_section}).}

\label{aper_comp}
\end{figure}

The dominant effect of this procedure is to produce a small
improvement in the achieved RMS scatter for the bright stars in the
sample.  Figure \ref{aper_comp} shows a comparison between the results
of using this procedure, and using only the $r_{\rm core}$ (smallest)
aperture.  We have used a simple constant multiplier to normalise each
frame to the photometric system of the master frame, via an
iterative $k \sigma$ clipped fit to remove any variable stars.

\subsection{Normalisation}
\label{poly_section}

The dominant effect of the atmosphere in ground-based differential
photometry is a time-variable shift in the photometric zero point of
each frame in the time-series.  This can result from the combination
of several effects, and is dominated by variations in transparency and
overall extinction (including the airmass-induced change in the
extinction seen on the frame).  Nightly zero point correction using
photometric standard star fields, as is commonly done for measuring
absolute photometry, is sufficient to reach the level of a few percent
down to $\sim 1 \%$.  Considerable progress can be made for the
purposes of differential photometry, especially over small fields of
view, by using non-variable stars in the field of interest to compute
zero point shifts for each frame in the time-series.

\begin{figure}
\centering
\includegraphics[angle=270,width=3in]{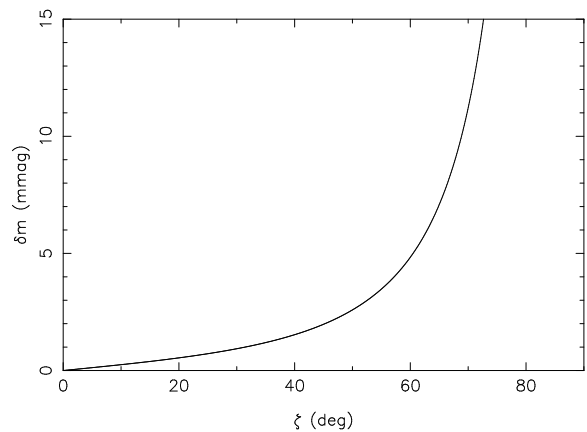}

\caption{Differential extinction across a $0.8\ {\rm deg}$ field as a
  function of zenith distance, for an assumed atmospheric extinction
  of $0.1\ {\rm mag\ airmass^{-1}}$.}

\label{ext_plot}
\end{figure}

For wide-field instruments such as the ones we are using, higher-order
effects start to become significant.  In particular, over a $\sim 0.8\
{\rm deg}$ diameter field (eg. INT+WFC or CTIO+Mosaic from
corner-to-corner), differential variations in airmass across the frame
are no longer negligible.  Assuming the approximation for the airmass
\begin{equation}
X \approx \sec \zeta
\end{equation}
where $\zeta$ is the zenith distance, and differentiating,
\begin{equation}
{\delta X \over{X}} = {\tan \zeta}\ {\delta \zeta}
\end{equation}
Substituting a typical value of $\zeta = 30^\circ$, $\delta X
\approx 0.009$.  For a typical $V$-band atmospheric extinction of
$0.1\ {\rm mag\ airmass^{-1}}$, this contribution is $\sim 0.9\ {\rm
  mmag}$, and becomes larger moving away from the zenith.  Figure
\ref{ext_plot} shows the difference in extinction across a $0.8\ {\rm
  deg}$ field as a function of zenith distance.

Since there are other slowly-varying effects as a function of position
on the frame (eg. some flatfielding problems, astrometric errors
inducing position-dependent loss of flux from the apertures, etc.) we
have opted for a generalised approach of fitting 2-D polynomials to
the magnitude residuals for each non-variable reference star on each
frame, rather than enforcing the particular airmass dependence for
atmospheric extinction (and in our experience this technique does
indeed give better results).  We have found a quadratic of the form: 
\begin{equation}
\Delta m(x,y) = c_0 + c_1 x + c_2 y + c_3 x y + c_4 x^2 + c_5 y^2
\label{quad_equation}
\end{equation}
to be sufficient for all our wide-field data thus far, where $x$ and
$y$ are the pixel coordinates (with the means $\bar{x}$ and $\bar{y}$
subtracted to give a zero-mean coordinate system, which improves the
stability of the least-squares solution), $c_i$ are the polynomial
coefficients (fit for each frame from a number of non-variable
reference stars) and $\Delta m(x,y)$ is the zero point offset at the
position $x,y$ on the frame.

Non-variable stars can be identified automatically by using the RMS of
the lightcurves to reject any variable sources.  We have found that it
is often possible to compute this directly from the uncorrected light
curve to obtain the initial fit of (\ref{quad_equation}), and the
refine the solution iteratively by rejecting the most variable stars
at each stage.  This technique selects $\ge 100$ non-variable bright
stars on each CCD of the mosaic for the Monitor data.

\begin{figure}
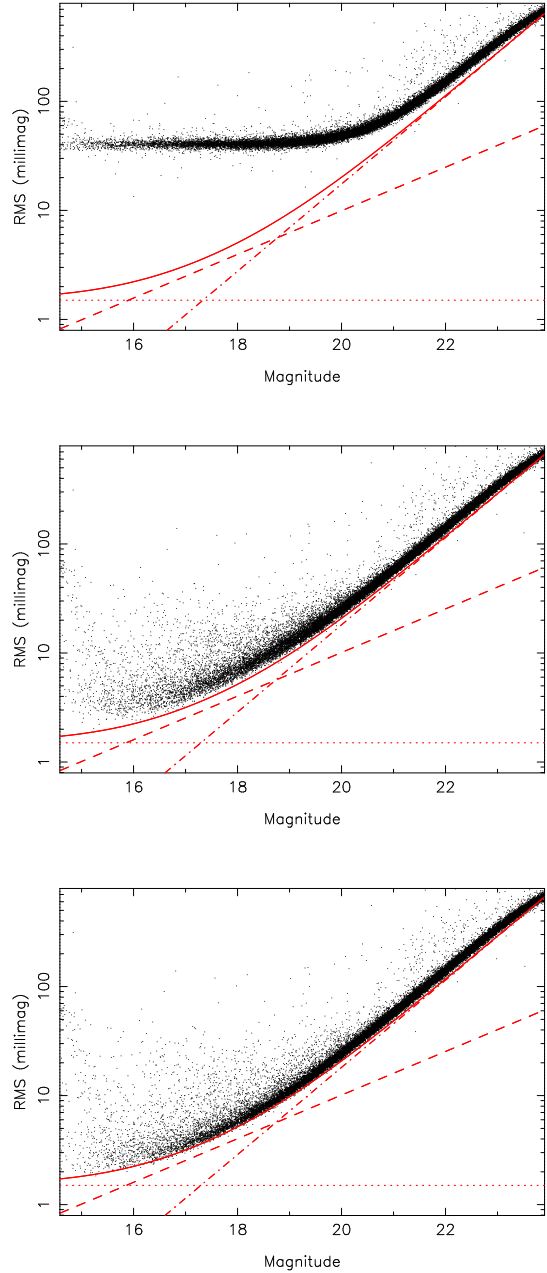

\centering
\includegraphics[angle=270,width=3in]{m50_rms_none.ps}
\includegraphics[angle=270,width=3in]{m50_rms_const.ps}
\includegraphics[angle=270,width=3in]{m50_rms_quad.ps}

\caption{Plots of RMS scatter as a function of magnitude for the
$i'$-band observations of M50, showing all objects of stellar
morphological classification.  The upper plot shows the results with
no zero point correction, the centre plot the effect of applying the
zero-order correction only, and the lower plot shows the full
quadratic correction.  Lines as Figure \ref{cent_comp}.}

\label{m50_rms_comp}
\end{figure}

Figure \ref{m50_rms_comp} compares the effects of applying no zero
point correction, a simple zero point shift, and the full quadratic
fit, for our CTIO-4m+Mosaic M50 data.  The best precision reached was
$\sim 35\ {\rm mmag}$ for the first case, $3\ {\rm mmag}$ with the
zero point shifts, and $2\ {\rm mmag}$ with the quadratic fit.

\subsection{Atmospheric scintillation}

Scintillation provides a fundamental limit to the noise performance
which can be reached in ground-based photometry.  Conventional results
for the scintillation level have typically assumed that one star is
observed at a time, and we might expect that some of the scintillation
would be cancelled out in CCD photometry due to the availability of
simultaneous observations of comparison stars.  However, \citet{rs98}
show that the typical coherence length is $\sim 12\ {\rm arcsec}$, so
over the fields of view we are considering, the single star result
should apply to a good approximation.  Therefore, we can adopt the
usual expression (see \citealt{rs98}) of:
\begin{equation}
{\sigma_{\rm scint} \over{F}} \approx 0.09 {X^{3/2}\over{D^{2/3}
    \sqrt{2T}}}\ \exp\left(-{h\over{h_0}}\right)
\end{equation}
where $\sigma_{\rm scint}$ is the RMS scintillation (in flux units),
$F$ is the object flux, $X$ is the airmass, $D$ is the telescope
aperture in centimetres, $T$ is the exposure time in seconds, $h$ is
the telescope altitude, and $h_0$ is a turbulence weighted atmospheric
altitude, taken here to be $h_0 = 8\ {\rm km}$.  For the INT+WFC
survey $i$-band observations this value is $0.44\ {\rm mmag}$, and for
CTIO+Mosaic $0.21\ {\rm mmag}$.  In both cases, scintillation is
negligible compared to the dominant noise sources in the data.  This
is nearly always the case for moderate exposure times on large
telescopes. 

\section{Implementation details}
\label{implement_section}

We present here some details of our actual implementation, as based on 
the discussion in \S \ref{dp_section}, for completeness.

The frame-to-frame astrometric transformations are computed
using a full astrometric model including radial distortions, by
performing an internal astrometric refinement.  A single data frame,
typically the one taken in best seeing and sky conditions, with a good
absolute astrometric solution (against 2MASS), is used as a reference.
The pipeline-generated object catalogue for this frame is used to
produce an astrometric reference catalogue, using the measured
positions for all bright, stellar sources (we use sources down to $2\
{\rm mag}$ below saturation).  The astrometric solution for each data
frame in the field is then refined against this reference.  The
internal accuracy after this procedure is typically $1/10\ {\rm
  pixel}$ or better.

We generate the master catalogue by stacking the $20$ data frames
taken in the best seeing and sky conditions, and use the standard
pipeline source detection and morphological classification software.
The classification software (see \citealt{mji06} for a more detailed
description) uses the flux of each object, measured in a series of
apertures of increasing radii: $r_{\rm core}/2$, $r_{\rm core}$,
$\sqrt{2}\ r_{\rm core}$, $2\ r_{\rm core}$ and $2 \sqrt{2}\ r_{\rm
core}$, where the default $r_{\rm core}$ is set approximately equal to
the FWHM of the stellar images.  By comparing these flux measures
(including also the peak height), the locus of stellar objects (which
all have approximately the same PSF and hence the same flux ratios
between apertures) is defined in planes of flux ratio as a function of
magnitude formed from several combinations of the measures.  This is
used to define a mean and standard deviation of the flux ratio for the
stars, as a function of magnitude, and a normalised statistic is
generated from this measuring how `stellar-like' each image is.  A
classification flag is subsequently derived by defining a boundary in
the statistic, and also factoring in the measured image ellipticities.

\section{Lightcurve production}
\label{lc_section}

We use a simple procedure for lightcurve production.  The first stage
is to convert all the flux measurements to magnitudes.  All of the
remaining stages of the processing are performed in magnitudes rather
than flux units for convenience.  Points with null or negative fluxes
(ie. below sky) are excluded from the lightcurves.  Each CCD of the
mosaic is processed separately (there are always enough stars to do
this in our fields of interest, otherwise we would have to use another
procedure).

The median and RMS flux of each object is calculated over all the
differential photometry measurements, using a robust MAD estimator
scaled to the equivalent Gaussian standard deviation (ie. $\sigma
\approx 1.48 \times {\rm MAD}$).  We apply the procedure of \S
\ref{poly_section} to fit and subtract a 2-D quadratic surface from
the residuals as a function of $x$ and $y$ coordinates on each frame.
In order to reduce contamination, the 2-D surface fits use inverse
variance weighting (using the RMS flux of each object calculated
earlier), and we exclude objects flagged as possible blends, saturated
datapoints, and all objects with non-stellar morphological
classifications.

We estimate expected per-datapoint photometric errors as the
quadrature sum of components from Poisson noise in the object counts,
Poisson noise in the sky, RMS of the sky background fit (multiplied by
the square root of the number of pixels in the aperture), and a
constant component of $\sim 1.5\ {\rm mmag}$ (as in Figure
\ref{m50_rms_comp}, for example) to account for systematic errors.
See \S \ref{noise_section} for a more detailed analysis of this last
component.

The lightcurves for each field are written into FITS binary tables in
multi-extension FITS files, with one extension per detector (this
convention is also used for the images, object catalogues and
differential photometry output).  These tables have one row per input
object from the {\em master catalogue}, and the lightcurve itself,
the photometric error on each lightcurve point and the heliocentric
Julian date of observation, are stored in columns of the table.  Our
lightcurve generation software, and this file format, have been
specifically designed to efficiently handle very large data-sets, for
example we have also successfully used them on data from the SuperWASP
transit search project \citep{p2006}.

At this stage, the data are ready for lightcurve analysis.  Our
analysis software, including period finding algorithms, an
implementation of the transit search algorithm of \citet{ai2004} and a
number of other programs, interface directly to the lightcurve FITS
files, and write their results out to additional columns in the files
for convenient storage.

Typically the full reduction of one week of data from the INT+WFC or
CTIO-4m+Mosaic takes $\sim 3\ {\rm days}$ including manual checking of
the pipeline results.  Often the most time-consuming stage of the
entire process is reading the data onto disk, which ranges from
relatively fast ($\sim 1\ {\rm day}$) using external IEE-1394 hard
disks (eg. for ESO WFI data), to very slow (up to $1\ {\rm week}$) for
DLT tapes.  We stress the increasing importance of this issue as data
rates from astronomical facilities continue to increase, and the
enormous savings in time and cost afforded by using internet transfers
(where possible) or efficient media such as external hard disks or
LTO-2 tapes.

\section{Noise properties}
\label{noise_section}

Lightcurves from ground-based transit surveys are invariably found
to show significant correlated, or `red' noise (see \citealt{pzq06}
for a very detailed discussion).  These correlations mean that,
averaging over $N$ data points, the error in the mean drops less
quickly than the `white' (uncorrelated) noise prediction:
\begin{equation}
\sigma_N = \sigma_0 / \sqrt{N}
\label{whitenoise_eqn}
\end{equation}
where $\sigma_N$ is the error in the mean of $N$ data points, and
$\sigma_0$ is the error in a single data point (where we have assumed,
for simplicity, that the uncertainties are equal for all the data
points).  Throughout this analysis, we assume a value of $N$
corresponding to $\sim 2.5\ {\rm hours}$, an appropriate timescale for
a hot Jupiter transiting a solar-like star, but also comparable to
timescales for eclipses in low-mass EBs.  We have tried to maintain
consistent notation with \citet{pzq06} throughout this Section.

The least-squares problem of finding the best-fitting box-shaped
transit model for a given lightcurve reduces to simply finding the
inverse variance weighted mean of the in-transit data points
(eg. \citealt{ai2004}), giving the transit depth if the mean of the
out-of-transit data points is subtracted.  In order to evaluate the
significance of a given detection, we use the detection statistic $Q$
of \citet{ai2004}, repeated here:
\begin{equation}
Q = \left(\sum_{i=1}^{N} {d_i\over{\sigma_i^2}}\right)^2 \left(\sum_{i=1}^{N} {1\over{\sigma_i^2}}\right)^{-1}
\label{q_eqn}
\end{equation}
where the summations run over all in-transit data-points $i$, $d_i =
f_i - \bar{f}$, the difference between the $i$th measured flux $f_i$
and the average flux $\bar{f}$ over all measurements, and $\sigma_i$
is the uncertainty on the $i$th flux measurement.

The presence of correlated noise in the lightcurves tends to give
larger values of $Q$ in the absence of transits.  Consequently, to
maintain a low false alarm rate, we must use a higher detection
threshold in $Q$, reducing sensitivity to shallow transits, or those
with few in-transit data points.  Furthermore, if the level of
correlated noise in each lightcurve is known, Eq. (\ref{q_eqn}) can be
modified to account for this in the transit detection process (see
\citealt{pzq06}).

We have examined the noise properties of our data using a method based
on that of \citet{pzq06}.  We present results based on the M50
lightcurves as a `best case' where we believe that our data
reduction is closest to optimal.  It should be noted that the
prescription we follow for evaluation of red noise will not work at
very faint magnitudes, where random noise sources dominate over the
correlated noise.  We have therefore analysed lightcurves of the
brightest non-saturated stars in our sample, where the effects of red
noise are much more significant.

\begin{figure}
\centering
\includegraphics[angle=270,width=3in]{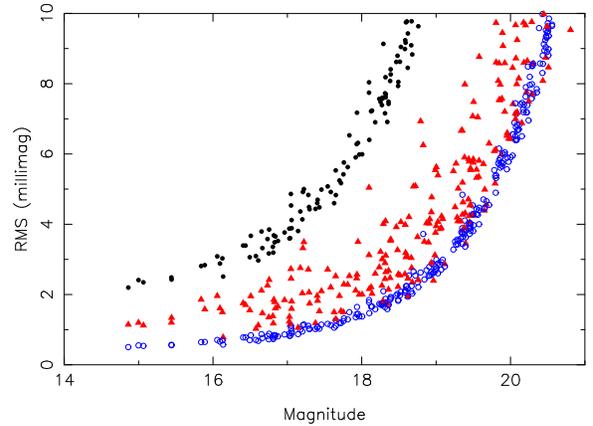}

\caption{Lightcurve RMS as a function of magnitude for a subset of M50
lightcurves not flagged as blended.  The symbols indicate the three
RMS measures: filled circles are values of $\sigma_0$, the RMS scatter
per data point, filled triangles are $\sigma_{2.5}$, the RMS scatter
of averages over $2.5\ {\rm hour}$ windows, and asterisks are
$\sigma_N$, the predicted RMS scatter over the $2.5\ {\rm hour}$
window assuming white noise.  The filled triangles lie between the
other symbols, indicating the presence of correlated noise at the
$\sim 1-1.5\ {\rm mmag}$ level over $2.5\ {\rm hours}$ for the
brightest stars, where the correlations dominate over random (white)
photometric noise.}

\label{m50_rednoise_1}
\end{figure}

Figure \ref{m50_rednoise_1} shows the RMS scatter as a function of
magnitude for a sample of lightcurves chosen to be approximately
`flat' (small reduced $\chi^2$), which should be noise-dominated.
We have calculated $\sigma_0$ and $\sigma_N$ from (\ref{whitenoise_eqn})
for $N = 19$, corresponding to $2.5\ {\rm hours}$ with the sampling of
these data, and compared $\sigma_N$ with $\sigma_{2.5}$ calculated as
the RMS of means over a $2.5\ {\rm hour}$ window moved along the
lightcurve.  This measures the correlated noise over a transit-length
window, and in general is larger than $\sigma_N$ if there are
correlations on this time-scale.  The results indicate that the level
of correlated noise on these time-scales is $\sim 1-1.5\ {\rm mmag}$
at the bright end.  Other teams have found instances of an increase in
the level of correlated noise at faint magnitudes, and Figure
\ref{m50_rednoise_1} shows that the same is true here for the majority
of the stars, where the $\sigma_{2.5}$ values never converge to the
$\sigma_N$ values.  Two likely causes of such effects are residuals in
the sky background determinations, and blending, both of which are
likely to affect faint stars close to sky more than bright stars.

\begin{figure}
\centering
\includegraphics[angle=270,width=3in]{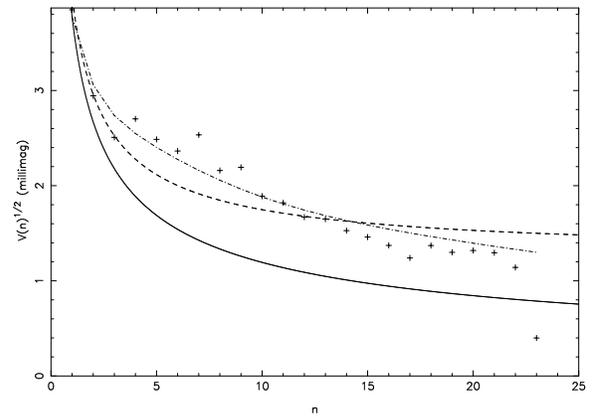}

\caption{The square root of $V(n)$ (ie. the standard deviation)
  plotted for a single `flat' lightcurve, with
  $\sigma_0 = 2.5\ {\rm mmag}$, $\sigma_{2.5} = 1.3\ {\rm mmag}$, and
  $\sigma_N = 0.6\ {\rm mmag}$.  The solid line shows the white noise
  prediction $V(n) = \sigma_w^2/n$, and the dashed line shows the
  fit of Eq. (\ref{var_equation}) to the data, with parameters
  $\sigma_w = 3.8\ {\rm mmag}$ and $\sigma_r = 1.3\ {\rm mmag}$.  The
  scatter (especially at large $n$) is caused by the limited number of
  $2.5\ {\rm hour}$ windows in the lightcurve containing these
  particular numbers $n$ of data points.  The dot-dashed line shows
  the predicted curve derived from the autocorrelation function of
  this lightcurve, using Eq. (\ref{var_acf_equation}).}

\label{m50_rednoise_single}
\end{figure}

In order to make a quantitative estimate of the level of correlations
in the noise, we have attempted to measure how rapidly the noise
`averages out' as a function of the number of data points observed
in-transit.  Figure \ref{m50_rednoise_single} shows the result for a
single `flat' lightcurve at the bright end of the RMS diagram ($I
\sim 15$).  In order to generate the diagram, a $\sim 2.5\ {\rm hour}$
window was moved over the data in $2\ {\rm minute}$ time intervals
(approximately $1/4$ of the sampling), counting the number $n$ of data
points lying in the interval, and recording the mean of the data
points.  We then computed $V(n)$ as the variance of the means at each
value of $n$ (where more than one mean was available).  For
uncorrelated (white) noise we expect $V(n) = \sigma_w^2/n$, where
$\sigma_w$ is the standard deviation of the white noise.  In general
there is an additional red noise component, which does not average out
as the number of data points is increased, ie.
\begin{equation}
V(n) = \sigma_r^2 + {\sigma_w^2\over{n}}
\label{var_equation}
\end{equation}
where $\sigma_r$ is the standard deviation of the red noise component.

\begin{figure}
\centering
\includegraphics[angle=270,width=3in]{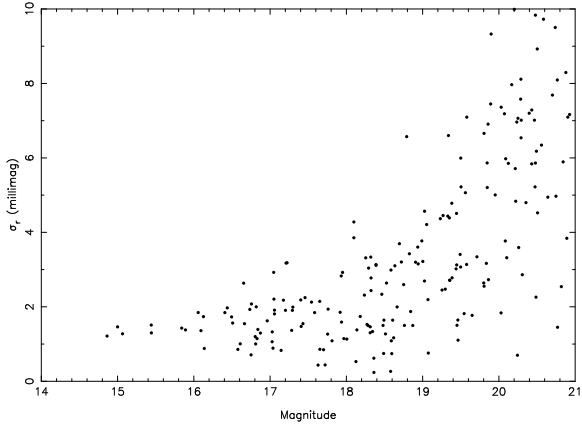}

\caption{Values of $\sigma_r$ from fitting Eq. (\ref{var_equation})
  plotted as a function of magnitude.  The upper envelope of derived
  values increases toward the faint end, which suggests that the red
  noise level increases for fainter stars.}

\label{m50_rednoise_2}
\end{figure}

Figure \ref{m50_rednoise_2} shows the values of $\sigma_r$ as a
function of magnitude for all the lightcurves in Figure
\ref{m50_rednoise_1}.  The upper envelope of derived values increases
toward the faint end, ie. the red noise level is higher at faint
magnitudes, as discussed earlier.  We note that the increased random
noise level at the faint end affects the determination of the values
of $\sigma_r$ (and $\sigma_0$), and hence introduces scatter as seen
in Figure \ref{m50_rednoise_2}.

\begin{figure}
\centering
\includegraphics[angle=270,width=3in]{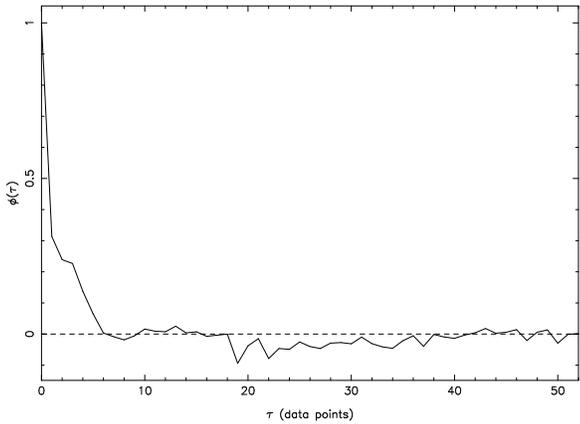}

\caption{Autocorrelation function of a `flat' M50 lightcurve,
  normalised to the zero-lag value ($\tau = 0$).  The sampling is
  approximately one data point every $6$ minutes, and the level of
  correlation is negligible for $\tau > 6\ {\rm data points}$.}

\label{m50_acf}
\end{figure}

An alternative method to investigate correlations among the
time-sampled data points is to compute the autocorrelation function.
Figure \ref{m50_acf} shows the autocorrelation function $\phi(\tau)$
of a representative `flat' M50 lightcurve, defined as:
\begin{equation}
\phi(\tau) = \sum_{n=1}^{N} \sum_{i=1}^{P_n} (m_{i,n} - \bar{m_n}) (m_{i+\tau,n} - \bar{m_n})
\end{equation}
where the outer sum is over nights of data $n$, and the inner sum over
data points within the night, up to the total $P_n$ taken in that
night.  $m_{i,n}$ is the magnitude of the star in measurement $i$ of
night $n$, and $\bar{m_n}$ is the mean magnitude of the star in
night $n$.  The summations were performed in this manner to avoid the
nightly gaps influencing the results for short time-scales.

The results indicate that the characteristic coherence timescale of
the correlations we see is $\sim 30\ {\rm minutes}$ (or $6$ data
points), which is typical of the `flat' lightcurves in the M50
data-set.

It is straightforward to show that the expected $V(n)$ can be
expressed in terms of the autocorrelation function as:
\begin{equation}
V(n) = {\sigma_w^2\over{n}} \left[1 + 2 \sum_{k=1}^{n-1} {n-k\over{n}}
 \ \phi(k)\right]
\label{var_acf_equation}
\end{equation}
This function is shown as the dot-dashed line Figure
\ref{m50_rednoise_single}, for an example lightcurve from the M50
data-set, and provides a better approximation to the observed
functional form for $n \la 10$ than the simple single-parameter
description of Eq. (\ref{var_equation}).  Note that
Eq. (\ref{var_acf_equation}) is not expected to exactly reproduce the
calculated $V(n)$ because for a given value of $n$, $V(n)$ counts only
2.5 hour windows containing $n$ data points, ie. for small $n$ the
function is dominated by the behaviour at the end of the night, or at
the end of observing windows interrupted by the weather, whereas the
ACF calculates the correlated noise over the entire lightcurve for all
$n$.

We find overall levels of `red noise' at the low end of the range
spanned by other surveys (eg. see \citealt{pzq06}, \citealt{s06}), of
$\sim 1.5\ {\rm mmag}$ at the bright end.  Since telescope time is at
a premium, we have only been able to use one observing 
strategy throughout, so it is difficult to quantify the factors
contributing to $\sigma_r$ from the present data-set.  However, since
our levels of red noise are comparable to the existing ground-based
surveys, we suggest that we may be obtaining close to the best
achievable performance for a ground-based survey over a $\sim 40\ {\rm
  arcmin}$ field using $2-4\ {\rm m}$ class telescopes, and that the
strategy of trying to keep the positions of the sources on the
detector as close as possible to constant, appears to be successful.
Nevertheless, it would be interesting to investigate the possibility
of using small random offsets to attempt to randomise the noise.

\section{Seeing-correlated effects}
\label{seeing_section}

We performed a search for correlations in the lightcurves with a
number of external parameters, including the image FWHM, sky level
(both globally and local to the sources), airmass, hour angle and
image morphology (major axis, ellipticity, position angle).  The
dominant effect was found to be seeing-correlated variations induced
by image blending.

Variations in the seeing cause an increase in the amount of
blended flux in the photometric apertures as the FWHM of the stellar
images increases, so therefore we expect to find a correlation between
the measured FWHM and the magnitude, for lightcurves of blended
objects.  This can be used both for flagging blended objects, and as
we shall see, for removing some of the variations induced by blending.

Our source detection software flags any objects where the deblending
algorithm (eg. \citealt{i85}) was invoked, and this flag is propagated
into the lightcurves to assist with identifying blended objects.  We
have found empirically that the flag is often set for objects which
do not exhibit any obvious blending effects in the lightcurves, since
a greater degree of overlap is required before the object lightcurve
becomes sufficiently contaminated.

\begin{figure}
\centering
\includegraphics[angle=270,width=3in]{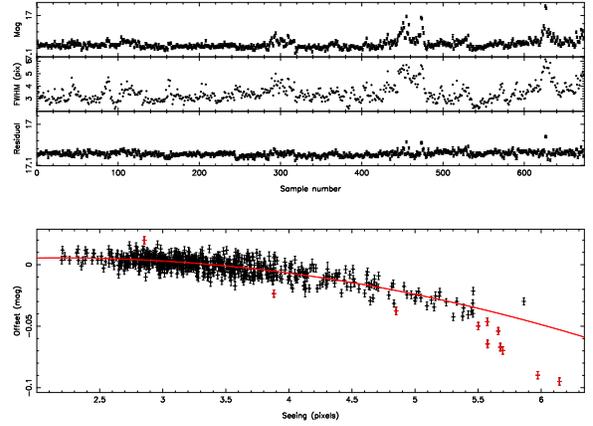}

\caption{Example of a lightcurve showing seeing correlations from our
CTIO M50 data.  The upper three panels show (from top to bottom) the
lightcurve, the seeing, and the residual after subtracting the
quadratic fit.  The lower panel shows the polynomial fit (solid line),
and the data plotted as error bars (points coloured red were excluded
by the iterative fitting procedure).  The statistic $b = 0.78$ for
this lightcurve, and the RMS was reduced from $7.2\ {\rm mmag}$ to
$5.3\ {\rm mmag}$ after subtracting the fit.  In this case the results
could be improved further by using a higher degree for the polynomial
(eg. a quartic).}

\label{blend_comp_1}
\end{figure}

\begin{figure}
\centering
\includegraphics[angle=270,width=3in]{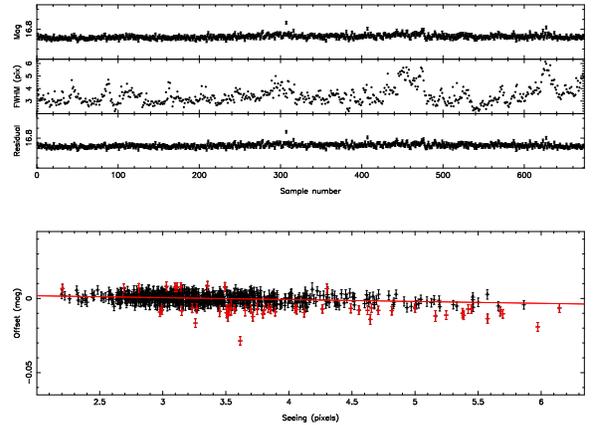}

\caption{Example of a lightcurve showing weak seeing correlations
from our CTIO M50 data.  Panels as Figure \ref{blend_comp_1}.  The 
statistic $b = 0.16$ for this lightcurve, and the RMS was reduced
from $3.6\ {\rm mmag}$ to $3.5\ {\rm mmag}$ after subtracting the
fit.}

\label{blend_comp_2}
\end{figure}

We therefore developed an empirical technique to characterise the
level of blending-induced effects in each lightcurve, by looking for
seeing-correlated shifts of the object from its median magnitude.
This is done by fitting a simple quadratic polynomial to the shift as
a function of the measured FWHM of the stellar images on the
corresponding frame.  Some examples are shown in Figures
\ref{blend_comp_1} and \ref{blend_comp_2}.  We use the following
statistic to quantify the level of blending:
\begin{equation}
b = {\chi^2 - \chi_{\rm fit}^2 \over{\chi^2}}
\end{equation}
where $\chi^2$ is defined as
\begin{equation}
\chi^2 = \sum_i {\left(\sum m_i - \bar{m}\right)^2 \over{\sigma_i^2}}
\end{equation}
for lightcurve points $m_i$ with uncertainties $\sigma_i$, and
$\bar{m}$ is the median magnitude in the lightcurve.  $\chi^2_{\rm
  fit}$ is the same statistic measured with respect to the quadratic
model.  $b > 0$ implies that $\chi^2$ was improved by the model fit --
ie. increasing values of $b$ to the maximum $b = 1$ imply
progressively greater amounts of seeing correlation in the lightcurve,
or increasing levels of blending.

\begin{figure}
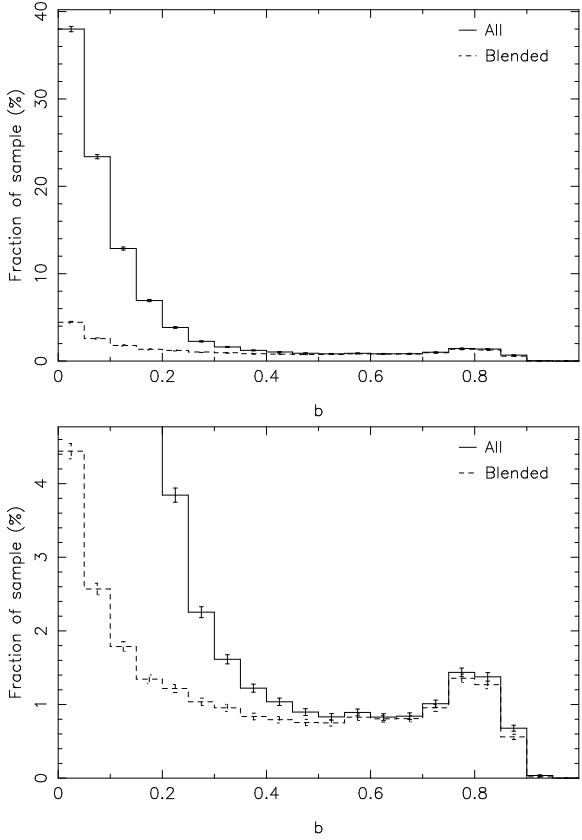

\centering
\includegraphics[angle=270,width=3in]{blend_1.ps}
\includegraphics[angle=270,width=3in]{blend_2.ps}

\caption{Histogram of the blend index $b$ for all lightcurves of
  stellar morphological classification in the CTIO M50 data.  The
  solid line includes all objects, and the dashed line only those
  objects flagged as blended by the source detection software.  The
  lower panel shows an expanded version of the upper panel.}

\label{blend_hist}
\end{figure}

Figure \ref{blend_hist} shows a histogram of the blend index,
indicating the presence of a peak at $b \sim 0.8$, corresponding to
objects showing clear seeing-correlated features due to blending, and
another peak at $b \sim 0$ corresponding to lightcurves without
seeing correlations.  The deblending flag from the source detection
software appears to work well for selecting lightcurves with no
blending, and hence no seeing correlation, but also flags relatively
large number of objects showing little or no seeing-correlated
behaviour, due to varying degrees of overlap.

\begin{figure}
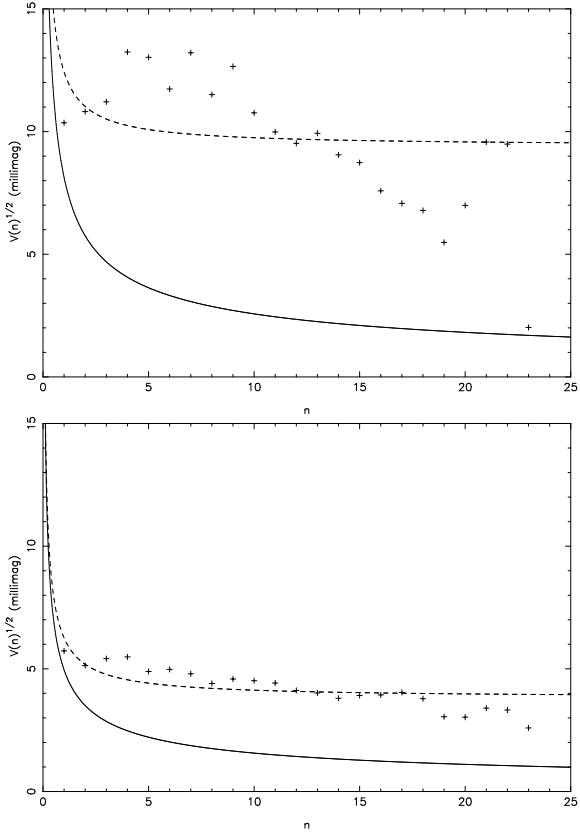

\centering
\includegraphics[angle=270,width=3in]{rednoise_1419_before.ps}
\includegraphics[angle=270,width=3in]{rednoise_1419_after.ps}

\caption{Plots as Figure \ref{m50_rednoise_single} for the object in
  Figure \ref{blend_comp_1} before (top) and after (bottom)
  subtracting the polynomial fit.  The value of $\sigma_r$ changed
  from $9.4$ to $3.8\ {\rm mmag}$, and $\sigma_w$ from $8.1$ to $5.0\
  {\rm mmag}$, indicating a significant reduction in the levels of
  white and correlated noise for this lightcurve.}

\label{blend_rednoise_single}
\end{figure}

\begin{figure}
\centering
\includegraphics[angle=270,width=3in]{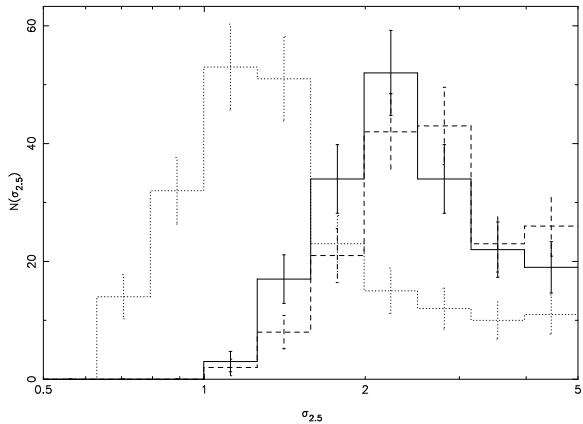}

\caption{Histograms of $\sigma_{2.5}$, the RMS scatter of averages
  over $2.5\ {\rm hour}$ windows, for all lightcurves flagged as
  possible blends on a single detector in the M50 data-set, before
  (dashed line) and after (solid line) the correction for
  seeing-correlated lightcurve features, showing the reduction in
  RMS resulting from the correction.  The dotted line shows the
  $1/\sqrt{N}$ prediction for white noise.}

\label{blend_rednoise}
\end{figure}

A natural progression from the analysis we have described is to
attempt to remove some of the seeing-correlated features in the
lightcurves by subtracting the fit.  Figures \ref{blend_comp_1} and 
\ref{blend_comp_2} show the results of doing this for two typical
lightcurves: one showing significant seeing correlated behaviour ($b
= 0.78$) and the other showing little seeing correlated behaviour ($b
= 0.16$).  In both cases, the procedure significantly reduces the
amount of seeing correlated features, and importantly, does not
introduce significant additional correlated features.  In both cases
the lightcurve RMS was reduced, as expected.  Figures
\ref{blend_rednoise_single} and \ref{blend_rednoise} show that this
corresponds to a reduction in the level of correlated noise as
measured in \S \ref{noise_section}.  We have used this simple approach
to produce a filter which can be optionally applied to our lightcurves
before embarking on transit searches and other similar analyses.

It is important to note that this approach to removing the effects of
blending, in reality, addresses the {\it symptom}, rather than the
{\it cause} of the problem.  Since aperture photometry (using multiple
apertures) is a simple approximation to full point spread function
fitting (hereafter, PSF-fitting), it is not surprising that heavily
overlapping images are not well-fit.

A conventional method for reducing the effects of image blending is to
move to PSF-fitting photometry (eg. \citealt{s87}), using analytical
or empirical PSFs, or a mixture of the two.  The use of PSF-fitting
brings with it a significant problem: that of accurately estimating
the PSF, which is particularly problematic over the wide fields of
view we are using due to the presence of significant PSF variations.

Difference image analysis (DIA; \citealt{al98}, \citealt{a2000}) is a
popular alternative, and is combined with aperture photometry, or even
PSF-fitting .  Briefly, in this method, the {\it master image} is
subtracted from each of the images in the time series.  The resulting
difference image should contain mostly noise, and only sources which
have varied in flux compared to the {\it master image} will remain.
In reality, the PSF varies from frame to frame on any real system,
which would leave residuals on the difference images, so it is
necessary to use an adaptive kernel \citep{al98}, which is convolved
with the master image to degrade the PSF to match each target image,
before subtraction.

DIA considerably simplifies the task of measuring photometry, since
the flux from blended stars is cancelled out if they do not vary (this
is nearly always the case) and therefore does not contribute to the
sums over the photometric apertures.  However, the method also
suffers from the problem of PSF estimation when computing the adaptive
kernel.  In most cases, PSF variations require a spatially-varying
kernel \citep{a2000} to produce good results and avoid leaving
residuals on the subtracted images for the non-variable stars.

Thus far, our attempts to use DIA have not produced superior results
to aperture photometry, although the work is still ongoing,
particularly in the ONC where extensive nebulosity limits the
photometric precision available from aperture photometry.  Particularly
in the case of our INT data, where the images have variable
ellipticities, we have found that the subtracted images contain 
significant residuals due to poor PSF matching, and these introduce
extra (correlated) noise into the lightcurves.  In these data, the
method does give some measurable improvement for blended stars, but
overall higher levels of correlated noise and occasional serious
lightcurve `glitches' in some objects.  We have therefore chosen to 
continue using aperture photometry, until we can resolve these
issues.

\section{The {\sc Sysrem} algorithm}
\label{sysrem_section}

This very popular method for finding (unknown) systematic effects
in time-series photometry was presented by \citet{t2005}.  The
{\sc Sysrem} algorithm resembles a generalised form of principal component
analysis (PCA), where the principal components are a set of
generalised `extinction' and `airmass' terms.  Mathematically, the
technique searches for the best two sets of coefficients $c_i$ and
$a_j$, to minimise the expression (in the notation of \citealt{t2005}):
\begin{equation}
S^2 = \sum_{i=1}^{N}\sum_{j=1}^{M} {(r_{ij} - c_ia_j)^2\over{\sigma_{ij}^2}}
\end{equation}
where the $N$ is the number of measurements in each lightcurve,
$M$ is the number of lightcurves, $r_{ij}$ is the residual
(mean-subtracted) flux of object $i$ on frame $j$, and $\sigma_{ij}$
is the corresponding uncertainty.  The products $c_i a_j$ can then be
subtracted from the lightcurves to remove this principal component,
and the technique repeated for subsequent components, deriving
progressively smaller corrections to the lightcurves.  Since the
coefficients are not constrained to be the actual extinction and
airmass, the technique also works for other forms of systematic
effect.

By examining the coefficients, it is possible to determine the origin
of the particular effect found by {\sc Sysrem}.  In particular the
terms $a_j$, representing the correction applied on each frame $j$ in
the time series, are often correlated with the parameters of the
images (eg. the seeing), pointing to the true cause of that particular
systematic effect.  We have therefore undertaken such an analysis to
find any residual effects in our data.

\begin{figure}
\centering
\includegraphics[angle=270,width=3in]{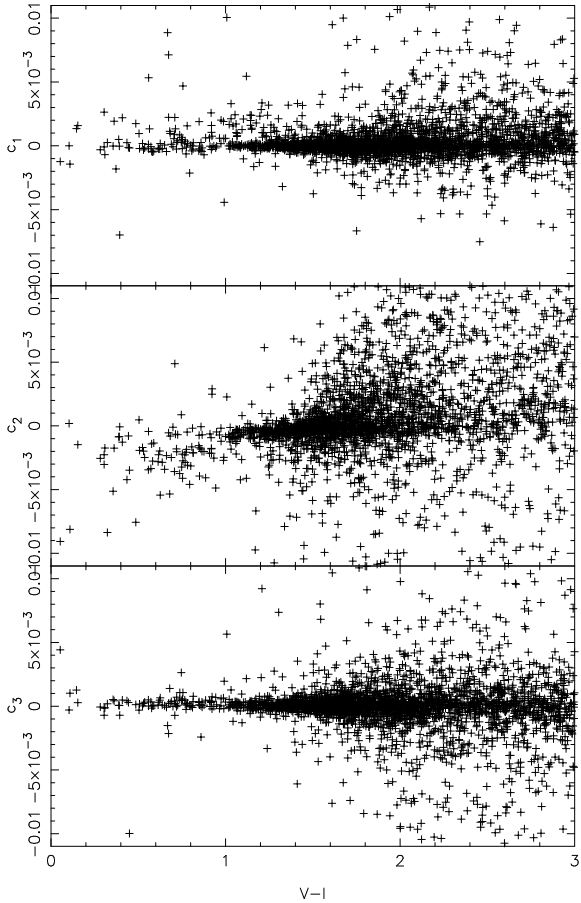}

\caption{Per-star coefficients $c_i$ for the first three {\sc Sysrem}
  components plotted as a function of $V-I$ colour.}

\label{sysrem_c}
\end{figure}

Figure \ref{sysrem_a} shows a plot of $a_j$ for the first three {\sc
  Sysrem} components, and for comparison, plots of several important
image parameters.  Figure \ref{sysrem_c} shows the coefficients $c_i$
for each star, plotted as a function of $V-I$ colour.  The first
component seems to show its largest values on a few non-photometric
nights, during periods of cloud.  There is no clear correlation with
$V-I$ colour.

\begin{figure*}
\centering
\includegraphics[angle=270,width=6in]{sysrem_a.ps}
\includegraphics[angle=270,width=6in]{trends.ps}

\caption{{\bf Top:} per-frame coefficients $a_j$ for the first three
  {\sc Sysrem} components plotted as a function of data point number.
  {\bf Bottom:} the corresponding values of (from the top): the zero
  order coefficient of the polynomial fit in \S \ref{poly_section}
  (mean extinction), image FWHM and offset of the centroid in the x
  coordinate. }

\label{sysrem_a}
\end{figure*}

The second component is clearly correlated with the image FWHM.  This
indicates that {\sc Sysrem} has found some residual effects of image
blending, not corrected by the method described in \S
\ref{seeing_section}.  This component is also mildly correlated with
$V-I$ colour (see Figure \ref{sysrem_c}), which indicates that a
wavelength-dependent effect (eg. extinction) has been detected.

\begin{figure}
\centering
\includegraphics[angle=270,width=3in]{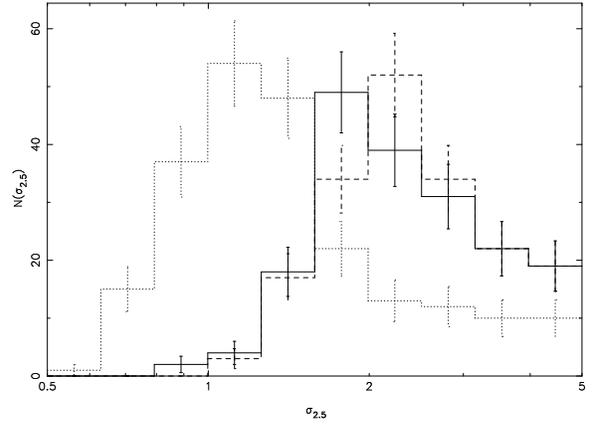}

\caption{Histograms of $\sigma_{2.5}$, the RMS scatter of averages
  over $2.5\ {\rm hour}$ windows, for all lightcurves on a single
  detector in the M50 data-set, before (dashed line) and after (solid
  line) removing the first two {\sc Sysrem} components.  The dotted line
  shows the $1/\sqrt{N}$ prediction for white noise.}

\label{sysrem_rednoise}
\end{figure}

The third component shows very little structure, and gives a
correction of very small amplitude ($\la 1\%$), with only one or two
frames having significantly non-zero values of $a$, and no correlation
with $V-I$ colour is apparent.  The effect of this component is
very small, and we conclude that for these data, the use of two {\sc
  Sysrem} components appears to be sufficient.  Figure
\ref{sysrem_rednoise} shows the result of subtracting off these two
components on the RMS over $2.5\ {\rm hour}$ intervals (approximately
the transit timescale).

The dotted line in Figure \ref{sysrem_rednoise} indicates that this
method has not detected all of the red noise sources present in the
data.  This conclusion is in agreement with the work of other authors
(Pont, private communication), and suggests that we still cannot fully 
describe the sources of correlated noise in time-series data using
the {\sc Sysrem} method.  This is most likely to arise for
effects which are not correlated between large samples of stars
(including the case where the effects are present in multiple stars,
but at different times).  We also note that some of the apparent `red
noise' could be due to very low-amplitude stellar variability.
\citet{tonry2005} find a very high occurrence of variability at the
few mmag level, which is included in our `red noise' estimates if it
occurs on a transit timescale.

It should be noted that we do {\it not} at present apply the
lightcurve corrections derived by {\sc Sysrem} (or the method of \S
\ref{seeing_section}) to our standard lightcurve output.  Instead, the
application of these filters is left to the user.  Specifically, they
have not been used for our rotation work (eg. \citealt{jmi06}) or for
visual transit searches, since at this level the systematics corrected
tend only to introduce (small numbers of) false positives, which can
be easily eliminated at the visual inspection stage, whereas the
subtraction of the {\sc Sysrem} corrections carries with it the risk
of introducing spurious variability from the residuals.

\section{Conclusions}
\label{conc_section}

We have developed a software pipeline for processing the high-cadence
time-series photometric data generated by the Monitor project, using
aperture photometry, to achieve RMS accuracy down to below $\sim 2\
{\rm mmag}$ at the bright end, typically with RMS $< 1 \%$ over $\sim
4\ {\rm mag}$ (eg. $13 < i < 17$ for the INT/WFC using $30\ {\rm s}$
exposures, $15.5 < i < 19$ for the CTIO-4m/Mosaic using $75\ {\rm s}$
exposures).  Our lightcurves are stored in a convenient FITS binary
table format, designed for efficient storage of multiple lightcurves,
and able to handle very large data-sets.

Noise properties of the data were investigated in \S
\ref{noise_section}, finding correlated (`red') noise at the level
of $\sim 1-1.5\ {\rm mmag}$ over a $2.5\ {\rm hour}$ transit-length
timescale.  These effects are important for transit searches since
they reduce the effective signal to noise ratio of the transit
detection statistic (here $Q$ as defined by \citealt{ai2004}), thus
leading to reduced sensitivity to low-amplitude transits and those
with few measured in-transit data points.  \citet{pzq06} examined the
effect of the level of correlated noise on the yield of Hot Jupiter
detections, finding that a level of $2\ {\rm mmag}$ gave a yield of
$\sim$ half the value for no correlated noise, as compared to $5\ {\rm
  mmag}$ for example, where the yield was $1/10$.  Therefore, we
conclude that the effects of correlated noise on the yield of our
survey are acceptable at the present level, but nevertheless we will
continue to pursue avenues for improvement such as PSF-fitting
photometry.

We have investigated seeing-correlated systematic effects in our
lightcurves induced by image blending.  A simple blend index was
developed to quantify the level of these effects seen in a given
lightcurve, based on the $\chi^2$ of a polynomial fit to the
lightcurve magnitudes as a function of the measured image FWHM (used
as an estimate of the seeing).  Subtracting the fit was found to be an
effective method for the removal of these seeing correlations, in lieu
of the use of techniques to properly eliminate the effects of
blending, such as PSF-fitting photometry and difference image
analysis.

Finally, the {\sc Sysrem} algorithm of \citet{t2005} was applied to the
data, and the effect of each component examined, to look for further
systematic effects.  The removal of two components was found to be
sufficient, with the first component removing some systematic effects
mostly associated with what appear to be particularly poor-quality
frames, and the second removing a seeing-correlated effect, most
likely due to residual image blending.  The second component is also
mildly correlated with $V-I$ colour, suggesting that this effect has
some wavelength-dependence, and may be related to atmospheric
extinction.

\section*{Acknowledgments}

The Isaac Newton Telescope is operated on the island of La Palma by
the Isaac Newton Group in the Spanish Observatorio del Roque de los
Muchachos of the Instituto de Astrofisica de Canarias.  Based on
observations obtained at Cerro Tololo Inter-American Observatory, a
division of the National Optical Astronomy Observatories, which is
operated by the Association of Universities for Research in Astronomy,
Inc. under cooperative agreement with the National Science Foundation.
This publication makes use of data products from the Two Micron All
Sky Survey, which is a joint project of the University of
Massachusetts and the Infrared Processing and Analysis
Center/California Institute of Technology, funded by the National
Aeronautics and Space Administration and the National Science
Foundation.

JMI gratefully acknowledges the support of a PPARC studentship, and SA
the support of a PPARC postdoctoral fellowship.  We also thank
Fr\'{e}d\'{e}ric Pont for useful discussions of correlated noise, and
Richard Alexander, Dan Bramich and Patricia Verrier for assistance
with the observing.  Sections \ref{noise_section} and
\ref{seeing_section} are based in part on discussions from meetings of
the International team on transiting planets of the International
Space Science Institute (ISSI), University of Bern.

Finally, we would like to express our gratitude to the staff of both
observatories -- the Isaac Newton Group and Cerro Tololo
Inter-American Observatory -- for their support.

\appendix

\section{Photometric errors from mis-centred apertures}
\label{applac_deriv}

In order to derive a simple analytic expression, let us consider a
source with a Gaussian PSF, centred on the origin, with total flux
$F_0$ and standard deviation $\sigma$.  Suppose that we use an
aperture of width $R$ in the $x$-direction, but integrate out to
$\pm \infty$ in the $y$-direction.  The flux measured, if this
aperture is perfectly centred, is given by
\begin{equation}
F = {F_0\over{2 \pi \sigma^2}} \int_{-\infty}^{\infty} dy\ \int_{-R}^{R} dx\ {\rm e}^{-x^2/2\sigma^2} {\rm e}^{-y^2/2\sigma^2}
\end{equation}

Now consider the case where the aperture is displaced by $\Delta$ in
the $x$-direction.  This modifies the limits of the $x$-integral thus:
\begin{equation}
F = {F_0\over{2 \pi \sigma^2}} \int_{-\infty}^{\infty} dy\ \int_{-R-\Delta}^{R-\Delta} dx\ {\rm e}^{-x^2/2\sigma^2} {\rm e}^{-y^2/2\sigma^2}
\end{equation}

Differentiating with respect to the shift $\Delta$ yields:
\begin{eqnarray}
{\partial F\over{\partial \Delta}} &=& {F_0\over{2 \pi \sigma^2}} \int_{-\infty}^{\infty} dy\ {\rm e}^{-y^2/2\sigma^2} \left[{\rm e}^{-(R - \Delta)^2/2\sigma^2} - {\rm e}^{-(R + \Delta)^2/2\sigma^2}\right] \\
&=& {F_0\over{\sqrt{2 \pi \sigma^2}}} \left[{\rm e}^{-(R - \Delta)^2/2\sigma^2} - {\rm e}^{-(R + \Delta)^2/2\sigma^2}\right]
\end{eqnarray}

Simplifying gives:
\begin{equation}
{\partial F\over{\partial \Delta}} = {F_0\over{\sqrt{2 \pi \sigma^2}}}\ {\rm e}^{-(R^2+\Delta^2)/2\sigma^2} \left[{\rm e}^{R \Delta/\sigma^2} - {\rm e}^{-R \Delta/\sigma^2}\right]
\end{equation}

For small $\Delta$, $R \Delta/\sigma^2$ will also be small, so we can
expand the exponentials in the final bracket to first order in this
quantity:
\begin{equation}
{\rm e}^{R \Delta/\sigma^2} - {\rm e}^{-R \Delta/\sigma^2} \approx {2 R \Delta\over{\sigma^2}}
\end{equation}

Furthermore, since $\Delta \ll R$, we can also approximate:
\begin{equation}
R^2 + \Delta^2 = R^2 \left(1 + {\Delta^2\over{R^2}}\right) \approx R^2
\end{equation}

Hence:
\begin{equation}
{\partial F\over{\partial \Delta}} \approx {F_0\over{\sqrt{2 \pi \sigma^2}}} {2 R \Delta
    \over{\sigma^2}}\ {\rm e}^{-R^2/2\sigma^2}
\end{equation}

Therefore, for small offsets $\Delta$, the resulting fractional error
in the measured flux is:
\begin{equation}
{\delta F\over{F_0}} \approx {1\over{\sqrt{2\pi}}}\ {\Delta
  \over{\sigma}}\ {2 R \Delta \over{\sigma^2}}\ {\rm e}^{-R^2/2\sigma^2}
\end{equation}

The expression will be non-analytic for a circular aperture with
finite extent in the $y$-direction, but the method we have used gives
a simple scaling relation to obtain an order of magnitude estimate of
the effect of mis-centring.

\end{document}